# Unlocking the Potential of Metaverse in Innovative and Immersive Digital Health

Fatemeh Ebrahimzadeh, Ramin Safa[*]

Department of Computer Engineering, Ayandegan Institute of Higher Education, Tonekabon, Iran.

**Abstract**

The concept of Metaverse has attracted a lot of attention in various fields and one of its important applications is health and treatment. The Metaverse has enormous potential to transform healthcare by changing patient care, medical education, and the way teaching/learning and research are done. The purpose of this research is to provide an introduction to the basic concepts and fundamental technologies of the Metaverse. This paper examines the pros and cons of the Metaverse in healthcare context and analyzes its potential from the technology and AI perspective. In particular, the role of machine learning methods is discussed; We will explain how machine learning algorithms can be applied to the Metaverse generated data to gain better insights in healthcare applications. Additionally, we examine the future visions of the Metaverse in health delivery, by examining emerging technologies such as blockchain and also addressing privacy concerns. The findings of this study contribute to a deeper understanding of the applications of Metaverse in healthcare and its potential to revolutionize the delivery of medical services.

**Keywords**: Metaverse, Healthcare, Virtual Reality, Artificial Intelligence, Machine Learning.

## 1. Introduction

Various industries, including healthcare, have been influenced by Metaverse technologies that are advancing rapidly. The Metaverse is a three-dimensional (3D) virtual world on the Internet where people can use avatars to perform their daily activities, which may be true or not. Virtual Reality (VR) and Augmented Reality (AR) in healthcare applications are related to the unique qualities and potentials of Metaverse technology [1]. For instance, VR is being used for surgical simulations and medical practice exercises so that medical students can enhance their skills without risking harm to real patients. Moreover, telemedicine platforms have been developed for remote communication between patients and healthcare providers. It allows patients to get better healthcare services through remote consultations with experts [2]. Other activities planned within this context involve building more sophisticated virtual environments to diagnose diseases, developing personal therapy programs, and creating virtual spaces for collaborative medical studies. Such developments indicate that Metaverse technologies may significantly improve healthcare provision [3]. Expanding electronic health is critical as it offers opportunities to improve health delivery while focusing on patients' needs first [4, 5].

This paper will comprehensively explore the current applications and future possibilities of Metaverse technologies in healthcare. To begin with, it provides an overview of the Metaverse concept and its underlying technologies, highlighting key examples of their integration in different industries. It then moves into Metaverse applications in healthcare, namely virtual reality surgical training, telemedicine platforms, and immersive therapies for patient rehabilitation. Next, the paper goes deeper into various benefits and challenges associated with the implementations and some examples of successful cases. Lastly, it discusses what tomorrow holds for merging Metaverse technology through blockchain, privacy issues, and the overall improvement of healthcare services.

[*]*Corresponding author:* safa@aihe.ac.ir

## 2. Understanding the Metaverse

The Metaverse, an emerging technological paradigm, presents multifaceted opportunities for transforming healthcare [5]. This internet-based virtual world can be harnessed as a tool for medical education, telemedicine, remote patient-expert communications, monitoring, self-service health data management, and global collaborations in medicine [6]. Integrating Metaverse technologies like VR and AR has the great potential to improve healthcare quality, patient outcomes, and advancing medical knowledge [5-7].

The Metaverse is essentially a combination of physical and digital worlds where people interact via their personalized digital avatars [2]. Therefore, it creates a mixture of virtual reality and real-life experiences, which can be applied to various problems and opportunities within the healthcare arena. No other technology can match the Metaverse's simulation of practical situations globally, thus providing unique prospects for use in medicine [4].

### 2-1. Metaverse technologies

Underlying the Metaverse are several sophisticated technologies that make creating and immersion into virtual environments possible. These instruments provide user interaction and engagement with digital realities used in many industries [8].

The key technologies used in the Metaverse, as shown in Figure 1, include:

*Virtual Reality (VR)*: VR technology uses sensors and display devices to create entirely artificial, computer-generated 3D environments [9]. Users can interact and navigate this virtual world through specialized VR headsets and hand controllers, which enable an immersive experience typically delivered via smartphones, PCs, or game consoles.

*Augmented Reality (AR)*: AR superimposes virtual information or objects onto reality to merge the physical world and virtual elements. This is supported by mobile or dedicated AR cameras, which feed the augmented information and virtual objects onto the screen, turning it into a dynamic and interactive scenery against the real world [10].

*Mixed Reality (MR)*: MR incorporates both VR and AR to allow users to switch smoothly from one mode to another. MR produces accurate and stimulating information mixed with a real-life environment, thus creating a more comprehensive experience [11]. MR is a kind of Extended Reality (XR) that merges the virtual and real worlds [9].

*User Interaction*: This is through the usage of various methods of input that help establish communication between the user and the virtual/real world [1]. They use hand controls, motion sensors, and other advanced interfaces to interact with virtual elements in a Metaverse environment.

*Diverse Applications*: The entertainment sector, education sector, healthcare industry, design and manufacturing industry, architecture and construction industry, and defense are among the industries that widely adopted Metaverse technologies [12]. These applications provide users with different kinds of practical experiences, from entertainment to professional research and industrial purposes.

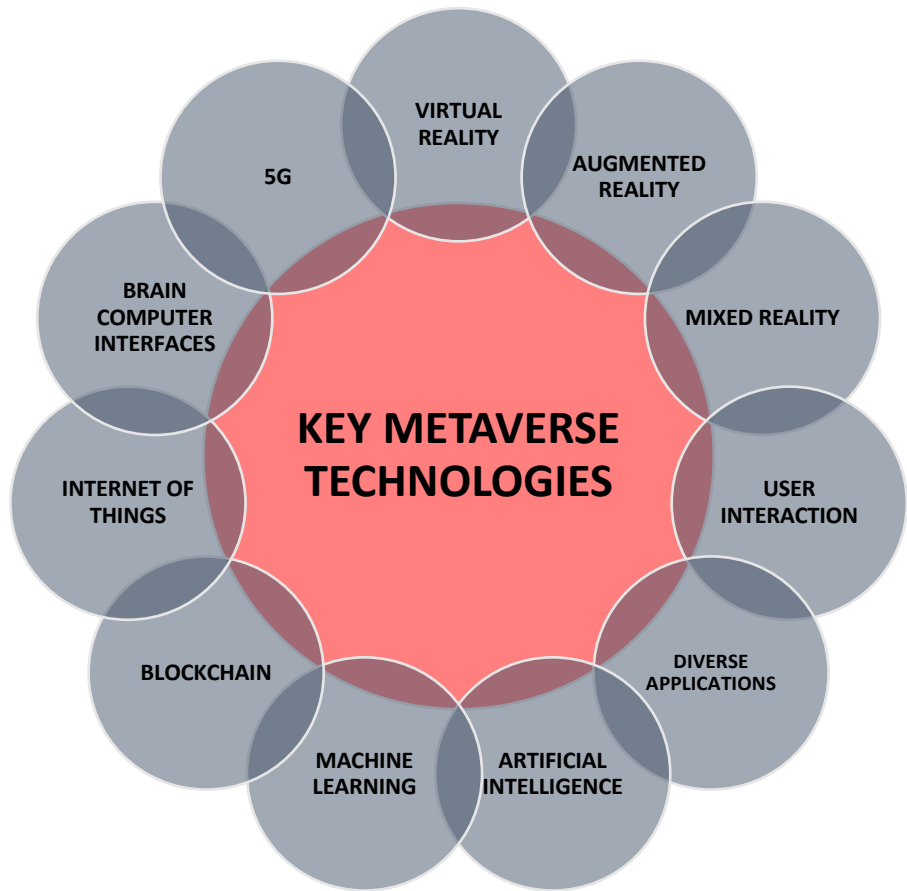

**Fig 1.** Common technologies used in the Metaverse.

## 3. The Metaverse in Healthcare

The integration of Metaverse technology in healthcare has a chance to boost the availability and efficacy of medical health services and offer novel therapeutic experiences for enhancing mental and physical well-being. Big tech firms are already studying the integration of these technologies into different healthcare applications.

### 3-1. Virtual Reality in the Metaverse

Devices such as specialized VR headsets and hand controllers serve as interfaces that allow users to create interactive computer-generated 3D environments that resemble simulation while being completely virtual [13]. On the other hand, AR combines virtual elements with the real world by projecting digital content and objects onto physical things using cameras of smartphones or dedicated AR devices [13]. One field where AR/VR has proved particularly beneficial is therapy [14-17]. Nowadays, this technology is used by psychologists and psychiatrists in the customization of individual patients' environments during aversion therapy, whereby patients encounter anxiety-causing situations within a controlled, safe virtual reality with close monitoring of their progress (Figure 2) [16]. Such personalized virtual therapy environments can be developed through mixed-reality technologies.

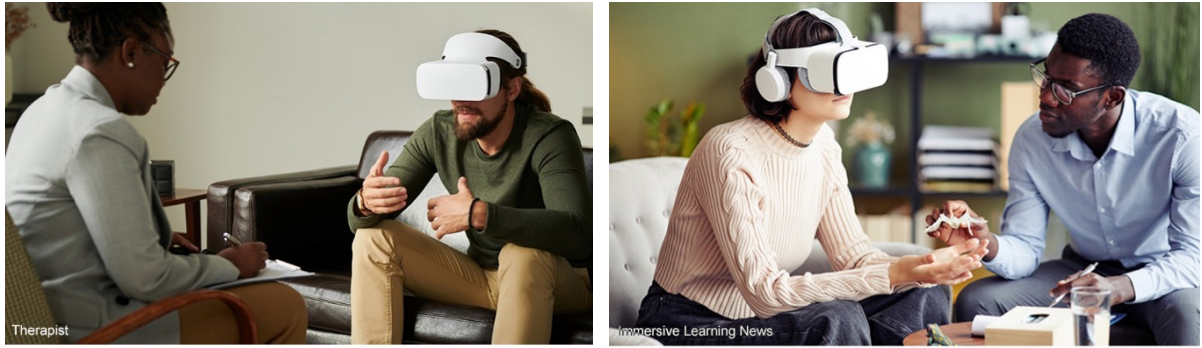

**Fig 2.** Using MR for virtual reality therapy.

### 3.2- Blockchain Technology

Blockchain technology's decentralized and secure nature makes it a promising solution for healthcare record storage and integration [18]. Blockchain's ability to enhance the efficiency of record storage, give individuals greater control over their data, and help address the limitations of existing systems positions it as a major driver for reshaping healthcare data management. One of the key benefits of using blockchain in healthcare is that such systems are decentralized, allowing the safekeeping of health records through personal e-wallets [19]. This decentralization method reduces the threats related to central repositories for data, which are prone to hacking regarding personal information [20]. By enabling people to manage their own medical information, blockchain-based health solutions can improve privacy and security while building trust between patients and doctors.

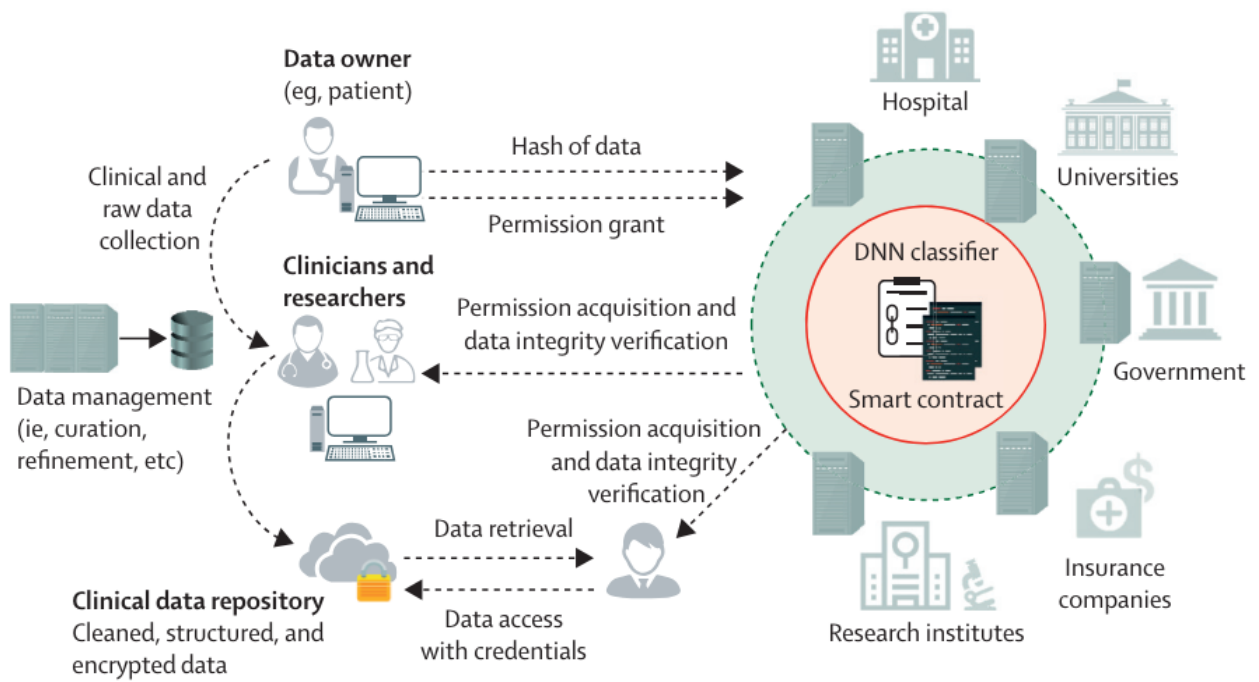

**Fig 3.** Blockchain-based Healthcare Data Management System between Multiple Stakeholders [21].

Figure 3 illustrates a new way of clinical data sharing advanced by blockchain. This framework allows patients to give some entities access to their medical information, thus increasing privacy, security, and control. In this case, a permission blockchain can be used by patients so that they can allow for fine-grained permissions specifying which data is accessed for what reason and by whom. It will also help in preventing unauthorized access to sensitive medical details. Moreover, smart contracts automate data sharing processes, facilitating workflows and promoting cooperation among healthcare providers, researchers, and other trusted parties.

Research on the use of blockchain in healthcare has remained ongoing, which helps unlock its full potential [19]. Integrating blockchain-based management within the evolving Metaverse ecosystem can potentially enable new transformative applications for public health.

### 3-3. General Applications of the Metaverse in Healthcare

The healthcare sector is full of potential to improve clinical outcomes highly, streamline medical operations, and enhance the patients' experience through integrating Metaverse technologies. The very first applications are going to be surgical simulations [7]. Thus, surgeons can practice complex procedures in virtual reality, experiment with new techniques, and refine their skills without any fear or risk of making mistakes [1, 4]. This could increase precision in surgery, decrease operating time, and result in better patient results.

With the Metaverse technology, clinicians could revolutionize how they handle medical imaging data for diagnostic purposes [22]. For example, CT scans, MRI images, and other modalities can be directly overlaid on a patient's body, rendering this information 3D, which is possible even when such a person moves [9]. Enhanced visualization provides for more accurate diagnoses and individualized treatment strategies.

Furthermore, Metaverse can enable innovative approaches to patient care management. Medical consultations over long distances and virtual clinics help more people access experts, especially those in underserved or remote areas [6]. Moreover, such engaging experiences as virtual reality-based rehabilitation programs permit patients to be actively involved in their recovery process, resulting in better general health and mental state.

Apart from the use of technology for medical treatments, there are other ways that Metaverse can improve healthcare provision. Using this platform in hospital resource allocation, workflows, and logistics through virtual simulations increases efficiency, reduces business costs, and enhances health delivery system performance overall [2-7]. How medicine is delivered or experienced will change as the Metaverse develops; it will also revolutionize how medical care is managed, leading to transformative patient outcomes and an equitable and sustainable healthcare system.

Various applications of the Metaverse in healthcare are illustrated in Figure 4. This figure includes patient monitoring, medical diagnosis, medical therapeutics, surgical procedures, and medical education.

These Metaverse-powered applications are built based on the digital twin concept as a key [23]. For instance, healthcare providers can make virtual biopsies, holographic imaging, gesture-based surgical aids, or even nanobot surgery, allowing virtual counseling, physiotherapy, and accelerated drug discovery. Integrating personalized, data-driven, immersive clinical experiences ultimately leads to better patient outcomes and more effective management of the clinical care infrastructure.

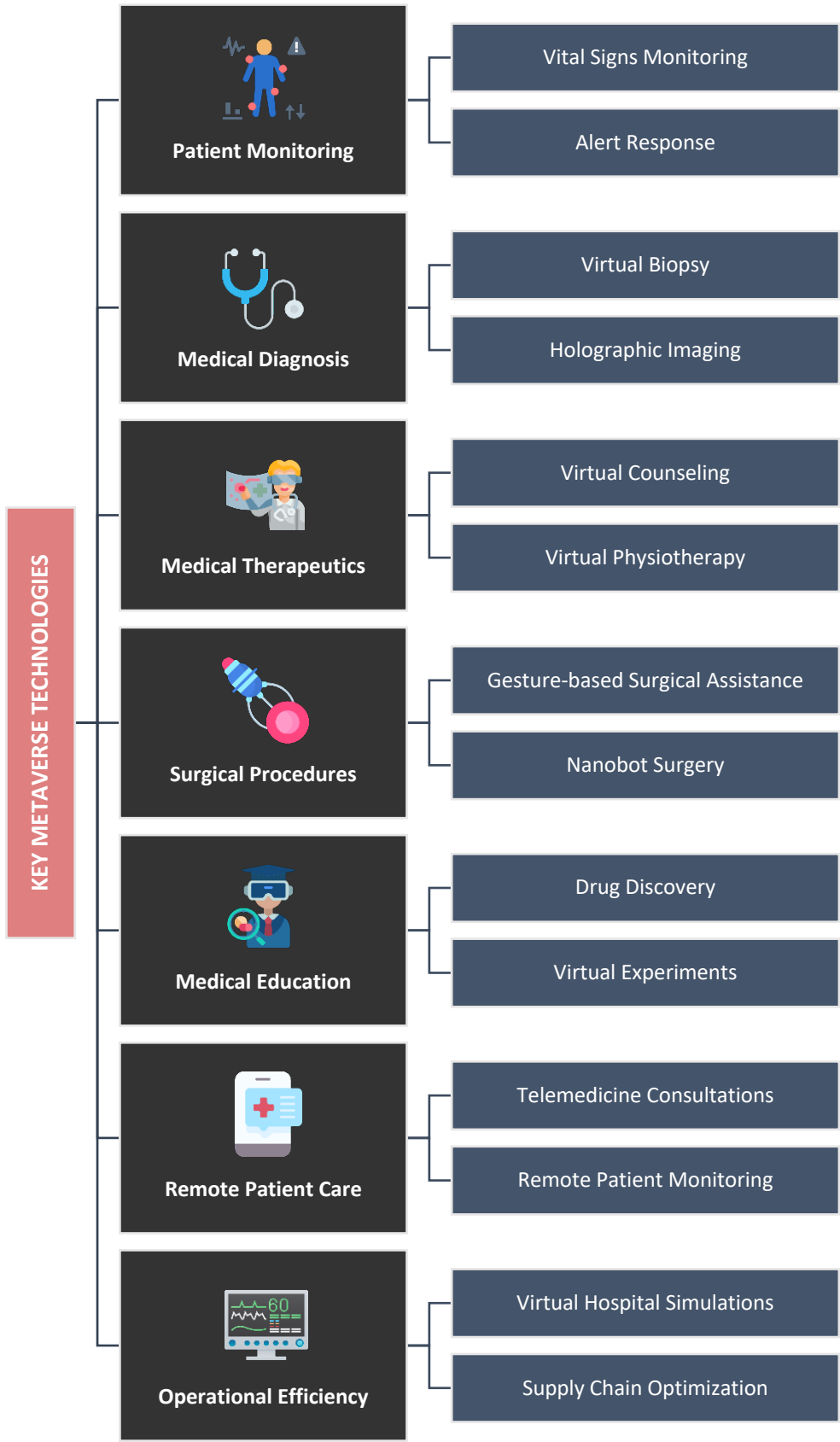

**Fig 4.** Metaverse Integration and the Future of Healthcare Delivery.

Healthcare could use Metaverse to combine services in various ways, and this has upsides and downsides (see Table 1) [1-7].

Table 1. Pros and Cons of Healthcare in the Metaverse.

| Pros | Cons |
|---|---|
| *Active Engagement*: Patients can access care in familiar environments, leading to increased interaction and activity. | *Access to Technology*: Not all patients and healthcare providers have access to the necessary technology or internet connectivity to effectively use Metaverse applications. |
| *Improved Communication*: Direct communication with healthcare providers through virtual platforms can enhance patient-provider interactions and provide opportunities for addressing patients' questions and needs. | *Privacy and Security Concerns*: Ensuring the security and privacy of patient data in a virtual environment is challenging and requires strong encryption and security measures. |
| *Comprehensive Information*: Access to educational resources and medical information via virtual platforms helps patients learn more about their condition and make informed decisions. | *Legal and Ethical Issues*: Using Metaverse technologies in healthcare raises questions about compliance with regulations, ethical considerations, and the need for new guidelines and standards. |
| *Informed Consent*: Patients must provide informed consent before engaging in virtual care, understanding the nature of the services, potential risks, and privacy implications. | *Implementation Costs*: The initial investment for developing and implementing Metaverse technologies can be high, posing a barrier for some healthcare providers. |
| *Enhanced Monitoring*: Metaverse-enabled remote monitoring and connected devices can provide continuous data collection, enabling more proactive and personalized care. | *Interoperability Challenges*: Integrating Metaverse technologies with existing healthcare systems and ensuring seamless data exchange can be complex. |
| *Immersive Therapy*: Virtual reality-based therapies can be used for pain management, phobia treatment, and rehabilitation, offering new avenues for patient care. | *Adoption and Acceptance*: Healthcare professionals and patients may hesitate to embrace Metaverse technologies, requiring extensive training and change management. |
| *Collaborative Care*: The Metaverse can facilitate virtual multidisciplinary team meetings, enabling healthcare providers to collaborate more effectively on patient cases. | *Regulatory Uncertainty*: Lack of clear regulatory frameworks and guidelines for using Metaverse technologies in healthcare can create uncertainty and slow adoption. |

### 3-4. Adoption of the Metaverse by Major Tech Companies

Major tech companies are already exploring integrating technologies into applications [24]. Meta Platforms[1] (formerly Facebook) is leveraging VR and AR technologies such as the Oculus Rift and Oculus Quest to develop social platforms that enable more immersive patient-doctor interactions and remote patient monitoring. Conversely, Google is utilizing its AR Core framework to create Android applications that overlay virtual elements onto the real-world environment, potentially aiding medical professionals in areas like diagnostic imaging. Similarly, Microsoft[2] is investing in Metaverse-based healthcare solutions, leveraging its HoloLens mixed reality technology and the Mixed Reality Toolkit framework for developing mixed reality applications. Also, Magic Leap[3], a pioneer in lightweight mixed reality technology and motion detection systems, provides Metaverse platforms for developing diverse augmented reality experiences in healthcare.

---

[1] https://about.meta.com/Metaverse
[2] https://microsoft.com/en-us/hololens/industry-healthcare
[3] https://magicleap.com

Likewise, Apple Vision Pro introduces an AR solution that has the potential to revolutionize surgical procedures. Surgeons with Vision Pro can directly make decisions on patients' health in the surgical field by projecting vital patient information and data into it. Rather than waiting for results, they can quickly analyze the situation and take appropriate actions. Consequently, there will be better outcomes in terms of precision.

According to the information on the Apple website[4], app developers are developing new apps for the Vision Pro that were not possible earlier, transforming areas such as clinical education, surgical planning, medical imaging, and behavioral health, among others.

Besides, game engine providers such as Unity[5] and Unreal Engine[6] contribute to the Metaverse ecosystem by offering tools and frameworks for developing VR and AR applications. These platforms facilitate the development of highly sophisticated applications that can be utilized in different fields within the healthcare industry.

## 4. Discussion, Future Direction, and Implications

Patient treatment, health outcomes, and satisfaction will improve when patients engage through Metaverse interaction with their specialists. Metaverse-supported virtual consultations and remote monitoring can expand healthcare access to underserved or rural areas and help keep an eye on chronic conditions [12]. It provides healthcare professionals with real-life training experiences through simulations and virtual environments that boost their skills and knowledge. Healthcare teams can use it for quick information sharing, case discussions, and informed decision-making, among others [3].

### 4-1. Emerging Opportunities and Challenges of Metaverse Integration in Healthcare

Clinical trials and research on Metaverse can offer fresh insights, accelerate the drug development process, and improve the overall quality and effectiveness of medical interventions [12]. However, several challenges in Metaverse healthcare have to be resolved. Healthcare systems operate within a complex regulatory landscape. Thus, integrating Metaverse technologies may call for new policies, guidelines, and legal frameworks to guarantee patients' privacy rights, data security, and ethical use of such technology [1-6]. Seamless data exchange and integration between Metaverse, existing healthcare information systems, and other relevant platforms are necessary to use Metaverse solutions effectively. Implementing Metaverse-based healthcare solutions requires robust, scalable, and reliable technological infrastructure, including high-speed internet connectivity, advanced computing resources, and user-friendly interfaces [32]. The ability of the Metaverse to gather sensitive health data raises significant privacy and security concerns that require strong measures and effective governance protocols to address overaccess and usage [25]. Also, widespread adoption and acceptance of Metaverse-based healthcare solutions will depend on overcoming user skepticism, addressing issues about the technology's usability and reliability, and providing comprehensive training and support to both healthcare professionals and patients.

---

[4] https://apple.com/newsroom/2024/03/apple-vision-pro-unlocks-new-opportunities-for-health-app-developers/
[5] https://blog.unity.com/topic/health-care
[6] https://unrealengine.com/en-US/spotlights/vr-medical-simulation-from-precision-os-trains-surgeons-five-times-faster

**4-2. Analyzing Metaverse Data and Machine Learning**

As we discussed earlier, Metaverse, besides being immersive and interactive, creates enormous data on user behavior, physiological responses, and interactions. When analyzed using machine learning approaches, such information has tremendous potential to transform healthcare. Machine learning algorithms can use Metaverse data to extract valuable insights from its huge collection [26-28]. For example, natural language processing methods can analyze text data from patient-provider interactions. Some particular domains for exploration include predictive analytics, virtual assistant chatbots, immersive therapy optimization, and clinical decision support systems. Figure 5 presents a machine learning-driven framework for bridging the Metaverse and healthcare. It shows the major steps of processing and analyzing Internet of Medical Things (IoMT) sensor data originating from the Metaverse devices.

Artificial intelligence models can recognize patterns and generate personalized risk assessments or predictive insights to enhance early intervention [18]. Social network analysis of Metaverse communities can also shed light on health-related behaviors, the spread of information, and the formation of peer support groups [29]. This knowledge can inform the design of more effective public health campaigns and interventions at the community level.

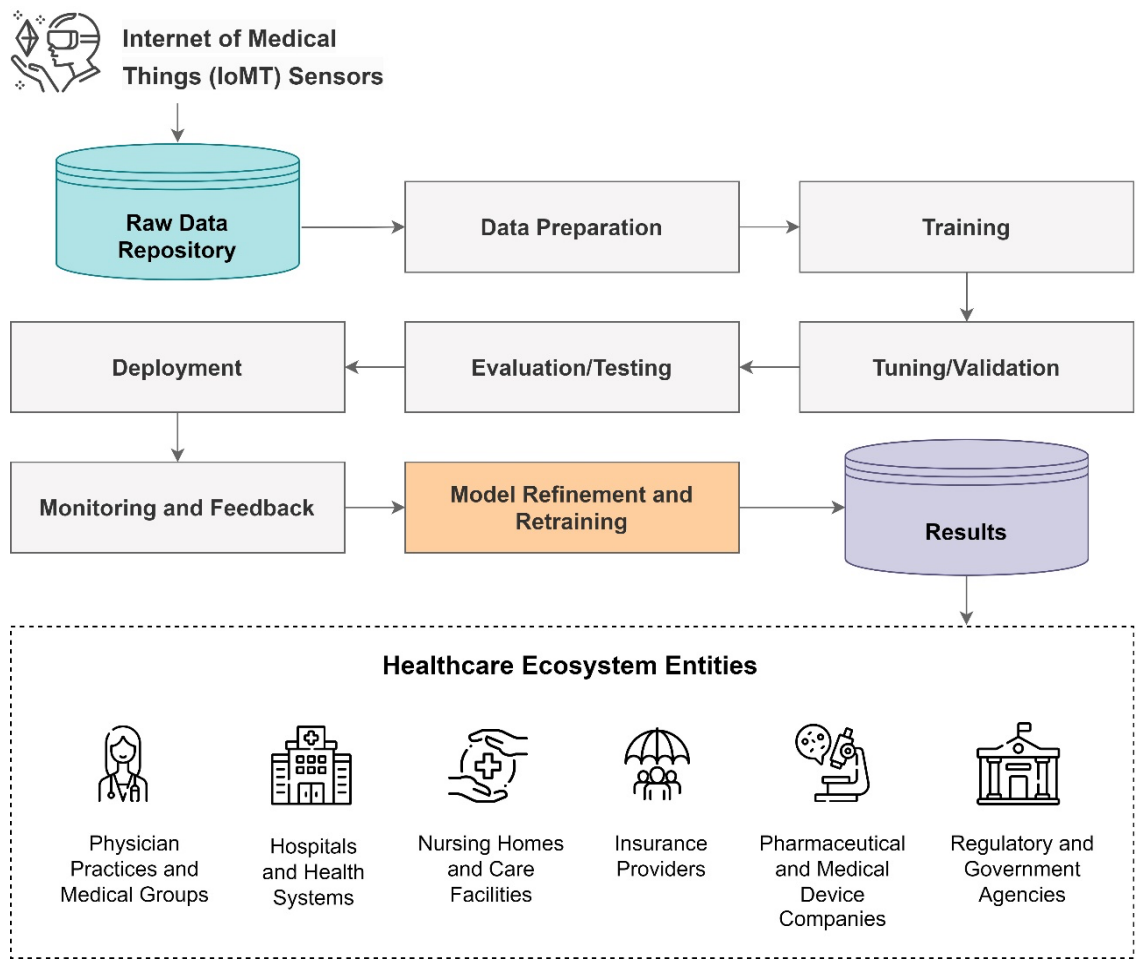

**Fig 5.** Bridging the Metaverse and Healthcare via Machine Learning.

Chatbots may be integrated within Metaverse to offer smart virtual patient assistance. Such bots can give them customized medical tips, symptom analyses, appointment scheduling, basic medical consultations based on users' profiles, and real-time interactions [30]. It means that machine learning algorithms could be trained using data obtained from Metaverse-based therapies (e.g., VR therapy) to enhance the personalization of these treatments, potentially leading to their effectiveness continuously increasing over time [31].

The Metaverse could become a transformative platform for healthcare by leveraging machine learning capabilities and data analytics to drive personalized, predictive, and preventive care solutions that ultimately lead to better patient outcomes and overall efficiency of healthcare systems.

### 4-3. Addressing Privacy Concerns in Metaverse-Based Healthcare Solutions

However, it can observe physiological responses and glean personal details about patients. The Metaverse brings up issues of privacy and security [23]. Innovative solutions like "clone cloud" or "private copy" can protect against malicious usage and data leakage. This paper underscores the importance of privacy in developing Metaverse healthcare approaches that engage the senses, communication, and behavioral aspects. Advancement in this area could uncover the transformative power of Metaverse in healthcare. In a virtual environment, hardware, software elements, and wearable devices are integrated into the Metaverse to ensure the confidentiality and consistency of sensitive information. Specifically, strong discourse norms and adoption strategies should be developed to address these challenges.

## 5. Conclusion

The employment of Metaverse technology in healthcare can greatly impact treatment delivery. From virtual reality surgical simulations to immersive therapies, Metaverse offers numerous possibilities for improving patient engagement, enhancing clinical results, and optimizing healthcare activities. It includes using machine learning approaches on data from the Metaverse to help personalize treatment plans, predict disease risks, and automate diagnosis. Incorporating these capabilities can improve patient outcomes and more efficient healthcare delivery. However, serious challenges such as data privacy, regulatory compliance, and technical infrastructure should also accompany these exciting opportunities. The transformative potential of the Metaverse in healthcare will need to unlock prioritizing robust security measures, developing clear ethical guidelines, and investing in reliable infrastructure.

Healthcare providers, researchers, and policymakers must collaborate as the Metaverse evolves to ensure its ethical and effective implementation. To implement it ethically and effectively, we need the contributions of healthcare providers, health informatics researchers, and policymakers, among other stakeholders, who will work with us. Our innovation is one way to address some concerns while capitalizing upon what this other world has in store for us so that all get accessible, personalized, and efficient health systems.

## Declaration

The authors declare they have used AI language models to provide editorial assistance with language clarity in preparing this manuscript.

# References

[1] G. Wang et al., "Development of Metaverse for intelligent healthcare," Nat. Mach. Intell., vol. 4, no. 11, pp. 922–929, 2022.

[2] G. Bansal et al., "Healthcare in Metaverse: A survey on current Metaverse applications in healthcare," IEEE Access, vol. 10, pp. 119914–119946, 2022.

[3] B. K. Wiederhold and G. Riva, "Metaverse creates new opportunities in healthcare," Annu. Rev. Cybertherapy Telemed., vol. 20, pp. 3–7, 2022.

[4] R. Chengoden et al., "Metaverse for healthcare: a survey on potential applications, challenges and future directions," IEEE Access, vol. 11, pp. 12765–12795, 2023.

[5] J. P. Venugopal, A. A. V. Subramanian, and J. Peatchimuthu, "The realm of Metaverse: A survey," Comput. Animat. Virtual Worlds, vol. 34, no. 5, p. e2150, 2023.

[6] A. S. Ahuja et al., "The digital Metaverse: Applications in artificial intelligence, medical education, and integrative health," Integr. Med. Res., vol. 12, no. 1, p. 100917, 2023.

[7] M. Sun et al., "The Metaverse in current digital medicine," Clin. eHealth, vol. 5, pp. 52–57, 2022.

[8] E. Dincelli and A. Yayla, "Immersive virtual reality in the age of the Metaverse: A hybrid-narrative review based on the technology affordance perspective," J. Strateg. Inf. Syst., vol. 31, no. 2, p. 101717, 2022.

[9] E. Dincelli and A. Yayla, "Immersive virtual reality in the age of the Metaverse: A hybrid-narrative review based on the technology affordance perspective," J. Strateg. Inf. Syst., vol. 31, no. 2, p. 101717, 2022.

[10] K. MacCallum and D. Parsons, "Teacher perspectives on mobile augmented reality: The potential of Metaverse for learning," presented at the World Conf. Mobile Contextual Learn., 2019.

[11] K. J. L. Nevelsteen, "Virtual world, defined from a technological perspective and applied to video games, mixed reality, and the Metaverse," Comput. Animat. Virtual Worlds, vol. 29, no. 1, p. e1752, 2018.

[12] P. Bhattacharya et al., "Towards future internet: The Metaverse perspective for diverse industrial applications," Mathematics, vol. 11, no. 4, p. 941, 2023.

[13] D.-I. D. Han, Y. Bergs, and N. Moorhouse, "Virtual reality consumer experience escapes: preparing for the Metaverse," Virtual Real., vol. 26, no. 4, pp. 1443–1458, 2022.

[14] M. B. Powers and P. M. G. Emmelkamp, "Virtual reality exposure therapy for anxiety disorders: A meta-analysis," J. Anxiety Disord., vol. 22, no. 3, pp. 561–569, 2008.

[15] P. M. G. Emmelkamp and K. Meyerbröker, "Virtual reality therapy in mental health," Annu. Rev. Clin. Psychol., vol. 17, pp. 495–519, 2021.

[16] M. Krijn et al., "Virtual reality exposure therapy of anxiety disorders: A review," Clin. Psychol. Rev., vol. 24, no. 3, pp. 259–281, 2004.

[17] I. Cortés-Pérez et al., "Virtual reality-based therapy improves balance and reduces fear of falling in patients with multiple sclerosis. a systematic review and meta-analysis of randomized controlled trials," J. NeuroEng. Rehabil., vol. 20, no. 1, p. 42, 2023.

[18] T. Huynh-The et al., "Blockchain for the Metaverse: A Review," Future Gener. Comput. Syst., vol. 143, pp. 401–419, 2023.

[19] X. Zhang, "Blockchain Technology based Metaverse Development Application," in 2023 IEEE 6th Information Technology, Networking, Electronic and Automation Control Conference (ITNEC), vol. 6, 2023.

[20] Y. Otoum et al., "Machine Learning in Metaverse Security: Current Solutions and Future Challenges," ACM Comput. Surv., vol. 56, no. 8, pp. 1–36, 2024.

[21] P. K. Ghosh et al., "Blockchain application in healthcare systems: A review," Systems, vol. 11, no. 1, p. 38, 2023.

[22] G. Wang, "Medical Imaging in Increasing Dimensions," Am. Sci., vol. 111, no. 5, pp. 294–301, 2023.

[23] M. Bordegoni and F. Ferrise, "Exploring the intersection of Metaverse, digital twins, and artificial intelligence in training and maintenance," J. Comput. Inf. Sci. Eng., vol. 23, no. 6, p. 060806, 2023.

[24] R. Gupta et al., "Are we ready for Metaverse adoption in the service industry? Theoretically exploring the barriers to successful adoption," J. Retail. Consum. Serv., vol. 79, p. 103882, 2024.

[25] R. Zhao et al., "Metaverse: Security and privacy concerns," J. Metaverse, vol. 3, no. 2, pp. 93–99, 2023.

[26] S. Danish et al., "Metaverse Applications in Bioinformatics: A Machine Learning Framework for the Discrimination of Anti-Cancer Peptides," Information, vol. 15, no. 1, p. 48, 2024.

[27] A. Qayyum et al., "Secure and robust machine learning for healthcare: A survey," IEEE Rev. Biomed. Eng., vol. 14, pp. 156–180, 2020.

[28] M. A. Rahman et al., "Towards machine learning driven self-guided virtual reality exposure therapy based on arousal state detection from multimodal data," in Int. Conf. Brain Inf., Cham: Springer, 2022.

[29] Y. A. Jeon, "Reading social media marketing messages as simulated self within a Metaverse: An analysis of gaze and social media engagement behaviors within a Metaverse platform," in 2022 IEEE Conf. Virtual Reality 3D User Interfaces Abstr. Workshops (VRW), 2022.

[30] A. Pourkeyvan, R. Safa, and A. Sorourkhah, "Harnessing the power of hugging face transformers for predicting mental health disorders in social networks," IEEE Access, vol. 12, pp. 28025–28035, 2024.

[31] O. Moztarzadeh et al., "Metaverse and healthcare: Machine learning-enabled digital twins of cancer," Bioengineering, vol. 10, no. 4, p. 455, 2023.